\def\year{2021}\relax
\documentclass[letterpaper,hyphens]{article} 
\usepackage{amstext}
\usepackage{booktabs}
\usepackage{hyperref}
\usepackage{comment}
\usepackage{aaai21}  
\usepackage{times}  
\usepackage{helvet} 
\usepackage{courier}  
\usepackage{graphicx} 
\usepackage{natbib}  
\usepackage{caption} 
\title{ “Two Many Cooks”: Understanding Dynamic  Human - Agent Team Communication and Perception Using Overcooked 2 }
\author{
    Andres Rosero, Faustina Dinh, Ewart J. de Visser, Tyler Shaw,  Elizabeth Phillips \\
}
\affiliations{
    George Mason University, Department of Psychology \\
    4400 University Drive, MSN 3F5\\
    Fairfax, VA 22030-4422\\
    arosero@gmu.edu,fdinh@gmu.edu
}
\begin{document}

\maketitle

\begin{abstract}
This paper describes a research study that aims to investigate changes in effective communication during human-AI collaboration with special attention to the perception of competence among team members and varying levels of task load placed on the team. We will also investigate differences between human-human teamwork and human-agent teamwork. Our project will measure differences in the communication quality, team perception and performance of a human actor playing a Commercial Off - The Shelf game (COTS) with either a human teammate or a simulated AI teammate under varying task load. We argue that the increased cognitive workload associated with increases task load will be negatively associated with team performance and have a negative impact on communication quality. In addition, we argue that positive team perceptions will have a positive impact on the communication quality between a user and teammate in both the human and AI teammate conditions. This project will offer more refined insights on Human - AI relationship dynamics in collaborative tasks by considering communication quality, team perception, and performance under increasing cognitive workload.
\end{abstract}

\section{Introduction}
 As a result of decades of innovation in the field of artificial intelligence, collaborative efforts between human and AI ‘agents’ have emerged across several fields. These human-agent teams (HATs) have been utilized to great success in fields like healthcare, military operations and education. Although human-agent teaming has become an expanding area of research and practice, there are still significant gaps in knowledge of how humans and autonomous agents work together in realistic, dynamic situations \cite{o2020human}. In particular, team processes like communication used when humans are working with autonomous agents may differ from humans working with other humans, and these differences may be especially apparent when teams encounter increasingly difficult and complex tasking and cognitive load. Research into human-human teams has shown that changes in cognitive workload inherent with more difficult tasking causes changes in communication amongst team members \cite{urban1996effects,xiao2003team}. Further, high performing human-human teams show patterns of communication that make their performance resilient to such changes in team workload that low performing teams do not \cite{converse1993shared}. When interacting in teams with autonomous agents, prior research has also revealed that human team members engage in fewer communications with autonomous teammates as compared to their human counterparts \cite{walliser2019team}. Such sub-optimal team communication processes can hinder the ability of the team to be resilient to changes in task and cognitive load and thus diminish performance of the team overall. In addition, when autonomous agents were framed and perceived as an interactive teammate and not simply an automated tool, participant communication with the autonomous increased significantly, and overall team performance was improved \cite{walliser2019team}. Taken together, these findings suggest that human perceptions of their artificial team members may be an important contributor to team processes and overall team performance. However, one notable gap in HAT research has been the influence of cognitive workload on team dynamics. Previous HAT research has focused on differences in perception in tasking where workload has been static, yet understanding team dynamics in situations with varying workload would have much greater ecological validity with real life human-agent teams completing tasks with varying difficulty across the lifespan of the team.

Thus, the purpose of this research is to investigate differences in team communication, team perception, and team performance between teams composed of humans and humans and artificial agents working together across varying levels of task load. In the following sections, we report progress on a research study where human actors will play a commercial off-the-shelf (COTS) cooperative video game, Overcooked 2, with either another human or a simulated autonomous teammate. 

\section{Background}
\subsection{Human-Agent Team Dynamics}
 There is inconsistency between human - human team perceptions and Human-agent team perceptions. Such was demonstrated in research led by Ashktorab in 2020, where although few differences were seen in performance between human-agent and human-human teams in a word guessing game, human participants perceived the agent teammates as less intelligent and likeable \cite{ashktorab2020human}. A research group led by O'Neill in 2020 conducted an extensive literature review of HATs and found that although human - human teams overall had higher levels of performance compared to HATs, higher levels of interdependence and greater levels of communication among HATs led to stronger coordination among members and greater team functioning.  

Extensive literature on human-human teams has shown that communication between collaborative members is important for team performance especially for distributing information and formulating strategies \cite{marlow2018does}. Consequently, if a comparison between human - human teams and human - agent teams is to be made, communication becomes a fundamental concern when evaluating HATs performance. Mou and Xu (2017) explored the initial stages of human communication with AI in comparison to human-human communications. The data revealed that the participants were less open and agreeable when interacting with AI than with other humans \cite{mou2017media}. These findings reiterate the necessity for further research on human-agent communication, specifically regarding how it may impact performance. 
\subsection{Input - Mediator - Output - Input Model}
The Input - Mediator - Output - Input (IMOI) model is an extension of the highly influential Input - Process - Output (IPO) model of team effectiveness posited by Hackman and Morris in 1975. The IMOI model expands upon the IPO by establishing that team processes are a complex system that involve interactions among inputs, outputs and mediators,not a causal chain from one process to the next \cite{ilgen2005teams}. For the purposes of this study, we focus on the IMOI in the context of virtual teams, a subfield that studies teams in which members are not co - located in the same environment and must operate virtually.

Explanations of the IMOI in the context of virtual teams have focused on the importance of communication as a key process towards the overall effectiveness of the virtual team \cite{marlow2017communication}. In 2017, Marlow and associates distinguished key attributes of communication which hold significance in team processes. Among these include communication frequency (how often team members communicate amongst themselves) and communication quality (an analysis of the clarity and effectiveness of interactions among team members). Due to the lack of pertinent direct and indirect interaction correlates such as non-verbal cues and body language \cite{dulebohn2017virtual}, the interplay between communication frequency and communication quality become integral to developing trust and cohesion among team members \cite{dulebohn2017virtual}. In addition, studies have noted that key aspects of the IMOI involve input processes such as team diversity (including teammate demographics and personality), mediators including task complexity and outcomes involving team performance and team satisfaction \cite{liu2010meta,dulebohn2017virtual,marlow2017communication}.

This study will leverage the IMOI virtual team framework to discern how team diversity (human v AI collaborator) affects processes of team effectiveness such as communication, perceptions of competence and performance while adjusting task complexity.

\begin{table*}[h]
    \centering
    \begin{tabular}{l l} \toprule 
  \textbf{Dependent Measures} & \textbf{Description}  \\\hline
    Communication Quality & Ratio: number of participant commands accepted by teammate in a level divided \\ 
    & by total participant commands \\ 
    Communication Frequency & Total instructions \\
    Team Perception & Combination of seven point team affect and team cohesion scales \cite{bishop2020chaopt} \\
    Cognitive Workload & NASA-TLX \cite{hart1988development} \\
        & Gas Tank Questionnaire \cite{monfort2018single} \\
    Task Load & Operationally defined from analysis of stimuli in each level based on a set list of criteria  \\
    Level score & End of level score provided by Overcooked 2 \\
    Positive Chef Actions (PCA) & The total sum of all actions that culminate in the creation of a dish in a given \\
    & Overcooked 2 level \cite{bishop2020chaopt} \\
    Chef Role Contribution (CRC) & The relative proportion of confederate 
    PCAs subtracted from the relative proportion of \\ 
    & participant PCAs. Positive CRCs indicate that the participant contributes more, negative\\
    & CRCs that the confederate contributes more \cite{bishop2020chaopt}\\\hline
    \end{tabular}
    \caption{List of Planned Project Measures and their descriptions}
    \label{tab:cap}
\end{table*}

\subsection{Commercial Off the Shelf (COTS) Games}
COTS games like Starcraft II \cite{vinyals2019grandmaster} and Hanabi \cite{eger2017intentional} have been utilized in an increasing number of human-agent teaming studies due to the gamification of communication and immersion necessary to obtain dedicated participation from participants. The division of labor and reliance on cooperation in game play to complete objectives makes it crucial for communication and interaction between teammates to occur - allowing researchers the opportunity to track changes in communication quality across tasking and game play. COTS games also allow researchers the ability to easily define performance within the context of the game. Additionally, a number of AI researchers have created approximations of COTS games to investigate models or human-agent coordination \cite{carroll2019utility, gao2020joint, song2019diversity, wu2021too}.

\begin{figure}[t]
    \centering
    \includegraphics[width=0.95\columnwidth]{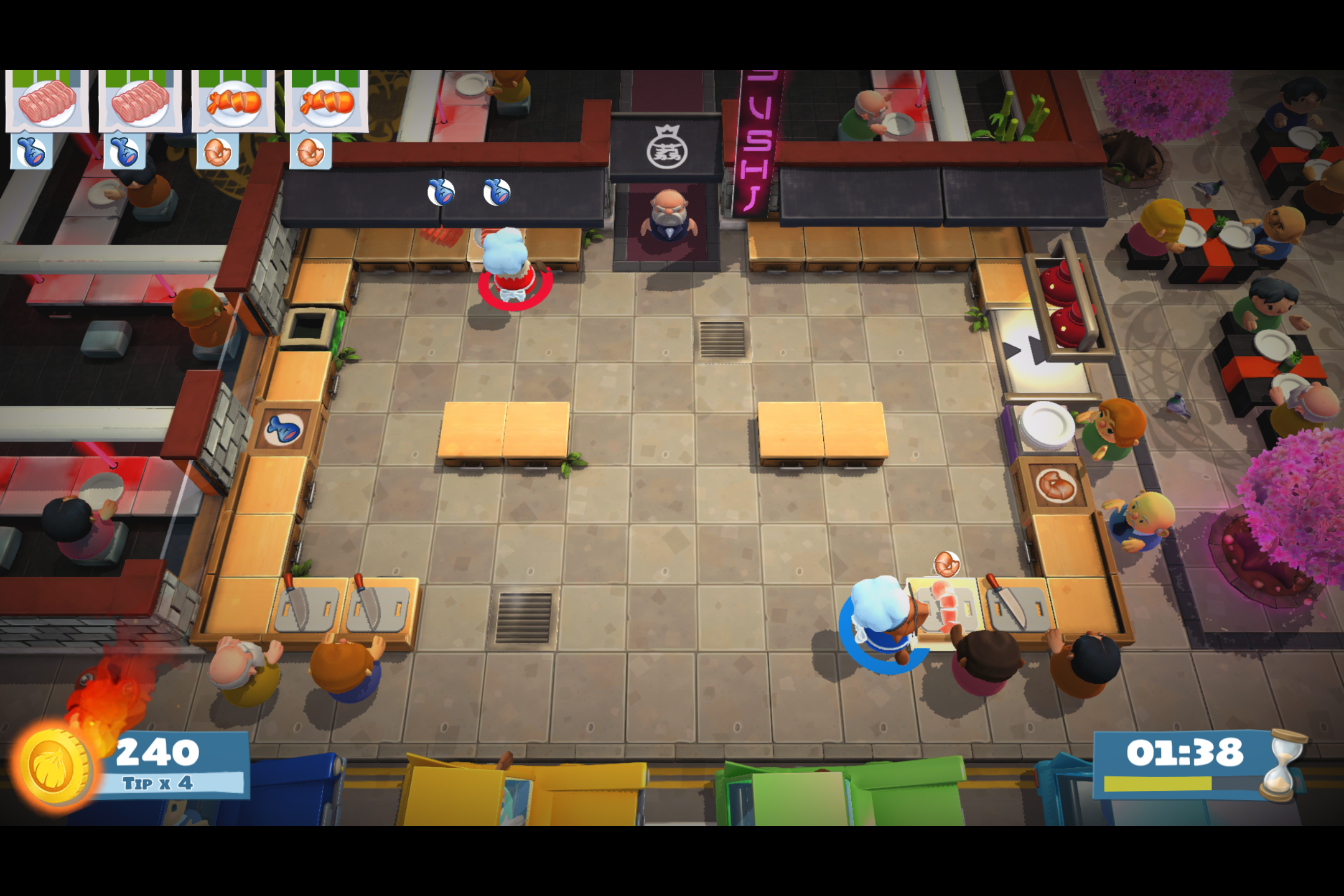}
    \caption{Screen capture of Overcooked 2 game play}
    \label{fig:task_stages}
    \end{figure}
    
Overcooked 2 (Figure \ref{tab:cap}) is a cooperative COTS game that relies on task asymmetry and frequent task reassignment in order to complete as many food orders as possible in a simulated restaurant within a predefined time limit. By requiring frequent division of labor to complete orders, Overcooked 2 encourages teamwork and communication among team members to complete orders promptly and gain high scores. 

The Cooking with Humans and Autonomy in Overcooked! 2 for studying Performance Teaming (CHAOPT) test-bed devised by a research group led by Bishop in 2020 determined that Overcooked 2 could be utilized as a viable platform for testing team processes such as communication, perception and performance \cite{bishop2020chaopt}. This work derived individual and team performance measures that could be gathered in-game and validated that these measures were sensitive to changes in level difficulty. 

Current research involving human - agent teams in COTS games has not discussed how changing task load affects team dynamics such as communication as a function of participant perception and its effect on team performance.  Task load measures the amount of stimuli and objectives present in a given task and influences the subjective cognitive workload of team members in a given task. In essence, as the amount of objectives and stimuli present in a given environment increase, there is an increase in the cognitive resource required to meet performance demands. Research has shown that increased workload has significant effects on communication among team members \cite{marlow2018does} and high functioning teams consistently show the ability to adapt their communication styles based on the difficulty of the task as well as the perception of team members confidence in the competence of other members to complete tasking \cite{wynne2018integrative}. By manipulating task load, this study will expand upon current COTS human - agent team research by examining dynamic team mechanics in varying task conditions, mirroring real life team objectives. We aim to see how human perceptions of the competence of a perceived AI teammate compared to a human teammate affect communication and performance as workload increases in accordance with the IMOI model of team effectiveness.

\subsection{Hypotheses}
Based on the review of previous research, the following hypotheses are proposed:
\begin{itemize}
    \item \textbf{H1}: Increased cognitive workload will be negatively associated in Overcooked team performance, such that as task load increases  cognitively overworked, Overcooked scores will suffer.
    \item \textbf{H2}: Positive team perceptions will have a positive impact on the communication quality between user and teammate in both the human and agent teammate condition.
    \item \textbf{H3}: Increased task load will negatively affect communication quality in both the human and agent conditions.
    \begin{itemize}
        \item H3.1: Communication quality will degrade at a greater level for the agent condition than human condition as a function of human perception towards teammate.
    \end{itemize}
    \item \textbf{H4}: Increased task load will negatively affect task performance in both human and agent teammate conditions.
    \begin{itemize}
        \item H4.1: Performance will be worse for the agent condition than human condition as a function of human perception towards teammate.
    \end{itemize}
\end{itemize}

\section{Proposed Methods}
\subsection{Participants}
We aim for a sample size of approximately 102 participants for this study in accordance with an a prior power analysis with a medium effect size. Participants will be recruited via Overcooked online forums on Reddit, Steam and Discord---targeting players who have completed the Overcooked 2 story mode and have access to online play on their computer and a working microphone. Only participants with Overcooked 2 experience will be recruited as a means of selecting players that could be capable of providing concise instruction to a confederate with knowledge of the game's mechanics.

\subsection{Measures}
Multiple existing subjective state scales and objective measures of process and performance will be employed before, during and after game play (See Table 1). 
We will use objective measures of team communication quality and frequency between teammates as well as objective measures of individual and team performance. Communication Quality is defined as the percentage of commands issued by the participant that were acknowledged and executed by the confederate. We will also include subjective measures of cognitive workload associated with changes in task load difficulty across game levels and subjective perceptions of the team itself (e.g., team affect, team cohesion). 

\subsection{Wizard of Oz paradigm} 
The Wizad of Oz (WoZ) paradigm is a common method for comparing perceptions between human-human and human - agent teams by controlling the capabilities of each teammate group \cite{musick2021happens}. The WoZ paradigm will be used to simulate agent activity with a human confederate simulating game play of an autonomous agent.By incorporating a WoZ paradigm, all reported differences in perception of competence across groups will be derived solely from the participant's belief in the identity of their teammate and not from the abilities of the teammate itself \cite{musick2021happens}.This study will implement a WoZ paradigm via online play - a human confederate will play as both a human and AI teammate. Communication from the confederate across both groups will be limited to only the emoticon system in Overcooked. The participant will be required to relay all commands verbally, and the confederate will respond with emoticons.

\subsection{Procedures}
Once recruited, participants will begin by completing an informed consent form. After consenting, participants will then be asked to view a short presentation detailing the tasks that participants will be expected to complete and instructions on how to communicate commands to the confederate teammate. Participants will interact with an experimenter guiding them through these materials remotely using Zoom, and Zoom will be used to audio and video record participant game play. Then, participants will be asked to complete a baseline measure of cognitive workload, the NASA-TLX, as well as a demographics survey containing questions pertaining to age and gender as well as experience with AI and with Overcooked 2. Materials will be prepared using the online study administration software, Qualtrics. Participants will be randomly assigned to two groups - a human teammate condition and an AI teammate condition.

Participants will enter into an online multiplayer game of Overcooked 2 and complete 6 experimental levels, three levels will be low task load levels and three will be high task load levels. For each participant, the first level will always be the tutorial level to introduce players to game play with the confederate and establish a baseline level of performance. The order of completion of the remaining 6 levels will be pseudo-randomized across participants using a Latin square design. After each level, participants will complete the the Gas Tank Questionnaire of subjective cognitive workload as well as the team perception measures. The experimenter will also video and audio record all game play in each level which will provide a recording of communication and individual and team performance. Once all levels are complete, participants will be asked to complete a one question manipulation check to discern if participants believed they were playing with an AI teammate as well as a post-task team perception survey, debriefed, and then compensated for their participation. All procedures will be reviewed by George Mason University's Institutional Review Board.

\section{Discussion and importance to HRI reseach}
This research is designed to enrich our understanding of human-agent
team processes in dynamic tasks with fluctuating task load. We plan to investigate the relationships between task and workload on team communication, perceptions of the team, and team performance and their comparisons to human-human teams. We hope to find significant changes in communication styles towards agent teammates as task loads increase as well as differences in communication and task performance relative to the Human - Human teams and influenced by participant perception of their teammate. 

There is some debate on the generalizability of HAT research to HRI domains because some claim that the physical embodiment of robots can alter their relationships with humans compared to disembodied AI counterparts \cite{o2020human}. We believe that the findings of this project will translate to Human-Robot interactions. As researchers like Nass and Scheutz have demonstrated, the dynamics by which Humans perceive non-human teammates with comparable trust and coordination as human counterparts is ubiquitous regardless of the form of the non-human teammate \cite{nass1996can,scheutz2017framework}. Further, artificial agents may be able to fluidly move into/out of and between "bodies" in the future \cite{williams2021deconstructed} and doing so will likely influence long term team interactions \cite{de2020towards} over time. By investigating team dynamics in adaptive task load in HATs, we gain an understanding of changes in human communication and perception that vary across task loads that we can utilize to apply to similar tasking with robotic agents. As both artificial agents and embodied robots become increasingly used as operational teammates \cite{phillips2011tools}, it will be important to understand the differences in how people engage in team process behaviors with these agents and consequently how we can encourage team process behaviors that will lead to good performance of the human-agent team overall. 

We envision the findings of this project to be a precursor for future HAT and HRI research. Projects of this kind would include robots assisting humans in increasingly difficult tasks and analyzing how human perception of the robot affects their communication towards the robot as well as the overall performance as tasking becomes more complex.
\bibliography{References}
\end{document}